\documentstyle[prl,aps,twocolumn,multicol,epsf]{revtex}
\tighten
\begin{document}
\twocolumn[\hsize\textwidth\columnwidth\hsize\csname
@twocolumnfalse\endcsname
\title{Radiation-reaction force on a particle plunging into
a black hole}
\author{Leor Barack}
\address{Department of Physics, Technion---Israel Institute of Technology,
32000 Haifa, Israel}
\author{Lior M. Burko}
\address{Theoretical Astrophysics,
California Institute of Technology, Pasadena, California 91125}
\date{\today}

\maketitle

\begin{abstract}
We calculate the self force 
acting on a scalar particle which is falling radially into a
Schwarzschild black hole. We treat
the particle's self-field as a linear perturbation over the fixed
Schwarzschild background. The force is calculated by numerically solving
the appropriate wave equation for each mode of the field in the time
domain, calculating its contribution to the self force, 
and summing over all modes using Ori's mode-sum regularization
prescription. The radial component of the force is attractive  
at large distances, and becomes repulsive as the particle approaches
the event horizon.
\newline
\newline
PACS number(s): 04.25-g, 04.70.-s, 04.70.Bw
\end{abstract}

\vspace{3ex}
]


The problem of finding the equations of motion for a particle in curved
spacetime is a long-standing open problem in General Relativity. Recently,
this problem has also become timely and extremely important. The planned 
Laser Interferometer Space Antenna (LISA) is expected to detect (among
other sources) the gravitational waves emitted from a compact object
orbiting a supermassive black hole (BH). Accurate templates,
which include also the radiation-reaction (RR) effects on the 
compact-object's
orbit, are essential for the detection of the signal. 

The traditional approach for calculation of the orbital evolution under
RR requires the calculation of the fluxes at infinity and through the
BH's event horizon (EH), of quantities which are constants of
motion in the absence of RR. Then one uses balance
arguments to relate these fluxes to the local quantities of the object 
\cite{poisson-leonard}.
However, such techniques typically fail,
because the evolution of the Carter constant, which is a non-additive
constant of motion, cannot be found by balance arguments \cite{hughes-00}. 

Several prescriptions to include the
RR effects in the orbital evolution have been suggested. Quinn and Wald
\cite{quinn-wald} and Mino, Sasaki, and Tanaka \cite{mino-sasaki-tanaka}
recently proposed general approaches for the calculation of self forces. 
However, it is not clear how to practically apply these approaches
for actual
computations, the greatest problem being the calculation of the non-local 
``tail'' contribution to the self force, which arises from the failure of
the Huygens principle in curved spacetime. More recently, Ori proposed a
practical approach for the calculation of the self force
\cite{ori,ori-unpublished}, which is based on decomposition
of the self force into modes, and on a mode-sum regularization
prescription (MSRP).
MSRP has been developed in full in Refs. \cite{barack-ori,barack} for a  
scalar particle in static spherically-symmetric
spacetimes, and has been applied for several non-trivial cases including a
static scalar charge outside a Schwarzschild black hole (SBH)  
\cite{burko-00}
and a scalar charge in circular orbit around a SBH 
\cite{burko-prl}. In addition, there is strong evidence that MSRP is
applicable also for electric-field RR
\cite{burko-00,burko-ajp}, and some evidence that it is applicable also 
for gravitational-field RR \cite{lousto}. 

MSRP has been directly applied until now only for stationary problems,
were the field was decomposed into Fourier-harmonic modes, and the
analysis was done in the frequency domain. 
This was easy to be done in \cite{burko-prl}
for the case of circular orbits around a SBH, because the
RR in that case admits a discrete spectrum. However, in general one faces
a time-dependent, evolutionary problem, and one expects the spectrum to
be continuous rather than discrete. In this paper, we apply MSRP for the
first time to a time-dependent, dynamical problem. 


We consider a pointlike massless particle of scalar charge $q$, moving
along a radial (timelike) geodesic outside a SBH of mass $M\gg |q|$, 
where the metric is 
$\,ds^2=-F(r)\,dt^2+F^{-1}(r)\,dr^2+r^2\,d\Omega^2$, $\,d\Omega^2$
being the metric on the unit 2-sphere, and $F(r)=1-2M/r$.  
Let the particle's worldline be represented by $x^{\mu}=
x^{\mu}_{\rm p}(\tau)$, with $\tau$ being the proper time 
along the geodesic. For inward radial geodesic
motion, to be considered here, we have (in Schwarzschild coordinates)
$\dot\theta_{\rm p}=\dot\varphi_{\rm p}=0$,
\begin{equation}\label{geodesic}
\dot r_{\rm p}=-\left[E^2-F(r_{\rm p})\right]^{1/2},
\quad\text{and}\quad \dot t_{\rm p}=E/F(r_{\rm p}),
\end{equation}
where a dot denotes $d/d\tau$, and
$E$ is the energy parameter
(which is a constant of motion in the absence of the self force).
The scalar field $\Phi$ coupled to the particle satisfies the
inhomogeneous wave equation
\begin{eqnarray}\label{KGequation}
\Box\Phi =
-4\pi q \int_{-\infty}^{\infty}
\delta^4\left[x^{\mu}-x_{\rm p}^{\mu}(\tau)\right](-g)^{-1/2}d\tau,
\end{eqnarray}
$g$ being the metric determinant, and $\Box$ denoting the covariant wave
operator. We next decompose 
$\Phi$ into modes 
\begin{equation}\label{decompose}
\Phi=\sum_{\ell=0}^{\infty}\phi^{\ell}=
2\pi q\sum_{\ell,m}Y_{\ell m}(\theta,\varphi)
\left[Y_{\ell m}^*(\theta_{\rm p},\varphi_{\rm
p})\,\frac{\psi^{\ell}}{r}\right],
\end{equation}
where $Y_{\ell m}(\theta,\varphi)$ are the standard scalar spherical
harmonics, an asterisk denotes complex conjugation, and $\psi^{\ell}=
\psi^{\ell}(r,t;r_{\rm p},t_{\rm p})$. By expanding the delta function
in Eq. (\ref{KGequation}) as 
$
\delta(\theta-\theta_{\rm p})\delta(\varphi-\varphi_{\rm p})
=\sum_{\ell,m}\sin{\theta}\,Y_{\ell m}(\theta,\varphi)
Y_{\ell m}^*(\theta_{\rm p},\varphi_{\rm p})
$
and using the orthogonality of the $Y_{\ell m}$,
we find that $\psi^{\ell}$ satisfies 
\begin{equation}\label{fieldequation}
\psi^{\ell}_{,uv}+V^{\ell}(r)\psi^{\ell}=S(r;r_{\rm p}).
\end{equation}
Here, $v\equiv t+r^*$ and $u\equiv t-r^*$
(with $r^*\equiv r+2M\ln[(r-2M)/2M]$) are the
ingoing and outgoing Eddington null coordinates, correspondingly, the
effective
potential is given by
$V^{\ell}(r)=(F/4)\left[\ell(\ell+1)r^{-2}+2Mr^{-3}\right]$,
and 
\begin{equation}\label{S}
S\equiv\frac{F}{2r}
\int_{-\infty}^{\infty}\delta(r-r_{\rm p})\delta(t-t_{\rm p})\,d\tau=
\frac{F^2}{2rE}\,\delta(r-r_{\rm p}).
\end{equation}
[In the last equality we use $d\tau=dt/\dot t=(F/E)dt$,
followed by integration over $t$.]
Finally, we express the modes $\phi^{\ell}$ in terms of the
($m$-independent) functions
$\psi^{\ell}$ by summing over the azimuthal numbers $m$ in Eq.\
(\ref{decompose}).
For radial motion we thus find that 
\begin{equation}\label{phi^l}
\phi^{\ell}=q\left(\ell+\frac{1}{2}\right)\frac{\psi^{\ell}}{r}.
\end{equation}

The {\em total} regularized self force (including both the local and the
tail parts) 
exerted on the scalar particle, $f^{\rm RR}_{\alpha}$, can be calculated
by \cite{barack-ori,barack,burko-amaldi}  
\begin{equation}\label{Ftotal}
f^{\rm
RR}_{\alpha}\equiv\sum_{\ell=0}^{\infty}\left[f_{\alpha}^{\ell\;\pm}
-A_{\alpha}^{\pm}\left(\ell+\frac{1}{2}\right)-B_{\alpha}\right]
\end{equation}
(evaluated on the particle's worldline), where 
$f^{\ell}_{\alpha}=q\phi^{\ell}_{,\alpha}$
is the (covariant) self-force contribution associated with the $\ell$-mode
of
the particle's self-field, and $A_{\alpha}^{\pm}$ and $B_{\alpha}$ are
regularization parameters, whose values are given by Eqs.\
(101) and (134) of Ref.\ \cite{barack}. For the radial geodesics 
considered here, these parameters take the form
\begin{equation}\label{A}
A_{r^*}^{\pm}=\mp\frac{q^2}{r^2}\,E,
\quad\quad
A_{t}^{\pm}=\pm\frac{q^2}{r^2}\,\dot{r},
\end{equation}
\begin{equation}\label{B}
B_{r^*}=-\frac{q^2}{2r^2}\left(2F-E^2\right),
\quad\quad
B_t=-\frac{q^2}{2r^2}\,\dot{r}\,E
\end{equation}
(with all quantities evaluated at $x^{\mu}=x^{\mu}_{\rm p}$).
One should use $f_{\alpha}^{\ell\;+},A_{\alpha}^+$ 
($f_{\alpha}^{\ell\;-},A_{\alpha}^-$) when one calculates the field's
gradient from
the $r\to r_{\rm p}^+$ ($r\to r_{\rm p}^-$) limit (in general 
$f_{\alpha}^{\ell\;+}\ne f_{\alpha}^{\ell\;-}$ \cite{barack-ori,barack}).
(Of course, the physical quantity $f^{\rm RR}_{\alpha}$ can be
derived from either of these two values, or from any of their linear
combinations.) In practice, we take below the $r\to r_{\rm
p}^-$ limit.

Thus, in practice, to derive the self force along any given radial
geodesic (parametrized by $E$), one should first solve Eq.\
(\ref{fieldequation}) for the various modes (with appropriately
chosen initial data---see below), then 
construct the quantities $f^{\ell}_{\alpha}$, and finally
sum over the regularized self-force modes using Eq.\ (\ref{Ftotal}).
This sum over modes is expected to convergence at least as $1/\ell$, as
the $O(1/\ell)$ term in the $1/\ell$ expansion of $f^{\ell}_{\alpha}$
vanishes \cite{barack-ori,barack}.

\begin{figure}
\input epsf
\centerline{\epsfxsize 8cm \epsfbox{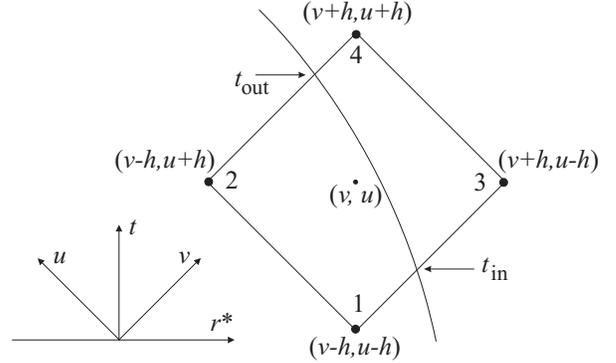}}
\caption{Numerical grid cell containing a section of the worldline.
(See the text for further explanation.)}
\label{unitcell}
\end{figure}

\begin{figure}
\input epsf
\epsfxsize=8.5cm
\centerline{ \epsfxsize 8.1cm
\epsfbox{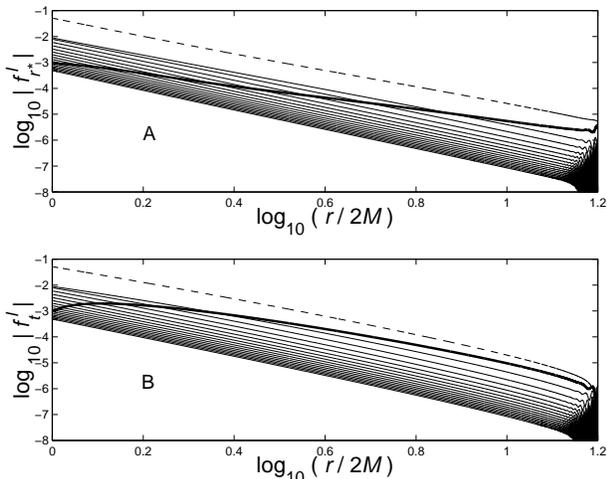}}
\caption{The individual modes of the regularized RR force
for a particle released from rest from $r^*_0=40M$ as a function of $r/M$.
Shown are the first 18 modes ($\ell=0..17$). The monopole ($\ell=0$) modes
are displayed by dashed lines, and the dipole ($\ell=1$) modes by thick
lines. Except for the $\ell=0,1$ modes, the modes' amplitudes decrease
monotonically with $\ell$.
Top panel (A): $f_{r^*}^{\ell}$. Bottom panel (B): $f_{t}^{\ell}$. }
\label{fig1}
\end{figure}

To solve for $\psi^{\ell}$, we integrate
Eq.\ (\ref{fieldequation}) numerically (in the time domain) on a
double-null grid.
This grid is spanned by $v$ and $u$, covering the entire exterior of the SBH
(with the EH approached at $u\to\infty$).
A characteristic initial-value problem for $\psi^{\ell}$ is set up by
specifying initial data on two null hypersurfaces $v=v_0$ and $u=u_0$,
taken to intersect at some point along the particle's
worldline. As initial data we take the exact solution corresponding to a
static particle held fixed at $r^*_0\equiv(v_0-u_0)/2$ 
\cite{burko-00}. This solution is not the actual initial field of the
geodesic particle considered here (which is unknown, in general). 
However, it does approximates the initial field if $r^*_0$ is
chosen to be a turning point of the geodesic 
(when such exists) (these initial data are inexact because the
acceleration of the geodesic particle at $t^*_0$ is
not the one of a static particle, 
although its position and velocity are), or---for a
marginally bounded particle (with $E=1$)---if
$r^*_0$ is taken large enough. The difference between the actual 
initial field and the static initial data results in the
occurrence of spurious waves superposed on the actual field;
however, one may expect such waves to quickly die off, unveiling
the intrinsic behavior of the field. Numerical experiments showed
that this is indeed the case: The spurious waves were found to
decay fast in all cases examined (see Fig. \ref{fig1}, e.g., for
$r^*_0=40M$).
For a marginally bounded particle it has been confirmed that the
field left after the spurious waves decay becomes independent of
$r^*_0$---indicating that one indeed extracts the actual physical
behavior. (In addition, we found that the larger $r^*_0$, the smaller the 
amplitude of the spurious waves, and the quicker they decay.)

To construct the difference scheme for the numerical integration we
use a method similar to that applied by Lousto and Price in
Ref.~\cite{lousto-price}.
We integrate the field equation (\ref{fieldequation}) over the unit
cell shown in Fig.~\ref{unitcell}, which is centered at $v,u$ and
whose sides are of length $2h$. Let $\psi_1\equiv\psi^{\ell}(v-h,u-h)$,
$\psi_2\equiv\psi^{\ell}(v-h,u+h)$,
$\psi_3\equiv\psi^{\ell}(v+h,u-h)$, and
$\psi_4\equiv\psi^{\ell}(v+h,u+h)$,
and suppose that $\psi_1$, $\psi_2$, and $\psi_3$ are already known, and
we wish to calculate $\psi_4$.
Integration over the $\psi^{\ell}_{,uv}$ term in Eq.\
(\ref{fieldequation})
yields (exactly) $\psi_1-\psi_2-\psi_3+\psi_4$.
Integration over the potential term yields 
$(\psi_1+\psi_4)[1+h^2V^{\ell}(r)]-(\psi_2+\psi_3)[1-h^2V^{\ell}(r)]+O(h^3)$. 
(Note that because $\psi^{\ell}$ is continuous across the worldline, the
integration of the potential term here is much simpler than in 
\cite{lousto-price}, where the metric perturbations were studied using the
Moncrief gauge, in which the wave function suffers a discontinuity across
the worldline.) Finally, integrating over the source $S$ (which is most
easily done by transforming to the $r,t$ coordinates, recalling that 
$\,dv\,du=2F^{-1}\,dr\,dt$), we obtain
$Z\equiv \int S\,dv\,du =0$
if the worldline does not cross the cell, or, if it does,
\begin{eqnarray}\label{Z}
Z=
E^{-1}\left[k(t_{\rm out})-k(t_{\rm in})\right]
(t_{\rm out}-t_{\rm in})+ O(h^3).
\end{eqnarray}
Here, $k(t)\equiv F[r_{\rm p}(t)]/r_{\rm p}(t)$,
and $t_{\rm in}$ ($t_{\rm out}$) is the $t$ value where the
worldline enters (leaves) the cell.
We can now extract the desired quantity $\psi_4$. To $O(h^2)$ 
we find
\begin{equation}\label{difference}
\psi_4=-\psi_1+\left[1-2h^2V^{\ell}(r)\right](\psi_2+\psi_3)+Z.
\end{equation}
Our code, which is second-order convergent, evolves the scalar field in a
straightforward marching. At each grid cell, the code does the following:
(i) it decides whether or not the given worldline crosses the cell;
(ii) if the worldline crosses the cell, it determines the point where
    it leaves it and calculates $t_{\rm out}$ (given $t_{\rm in}$) to
    $O(h^2)$;
(iii) it uses Eq.\ (\ref{difference}) to calculate the field $\psi^{\ell}$
at
    the cell's upper point; and
(iv) at grid cells containing a section of the worldline,
    it constructs the quantities $f^{\ell}_{r*}$ and $f^{\ell}_{t}$ by
    appropriately extrapolating the field gradients along the
    worldline (based on the already-derived values of the field at a few
    neighboring grid points).

We next present our results for a particle released from rest at
$r^*_0=40M$ (similar results are obtained also for other values of
$r^*_0$ and for the marginally-bound case). Figure \ref{fig1} displays the
behavior of the
$f_{r^*}^{\ell}$ (\ref{fig1}A) and $f_{t}^{\ell}$ (\ref{fig1}B)
components of the RR force. The $\ell=0$ components are
everywhere negative, whereas all the other modes ($\ell\ge 1$) are
everywhere positive. Figure \ref{fig1} also shows the decay of the
spurious waves. Clearly, for values of $r$ smaller than $25M$ 
they are already too small to be noticed.

\begin{figure}
\input epsf
\epsfxsize=8.5cm
\centerline{ \epsfxsize 7.5cm
\epsfbox{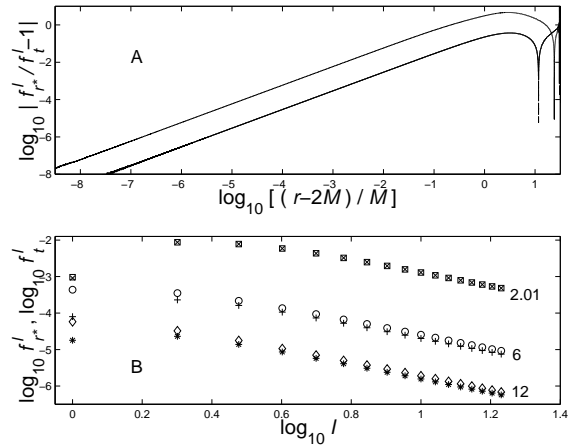}}
\caption{Top panel (A): The relative difference
$(f_{r^*}^{\ell}/f_{t}^{\ell})-1$ for
$\ell=0$ (upper) and  $\ell=1$ (lower) as functions of
$(r-2M)/M$.
Bottom panel (B): The behavior of the individual modes as a function of
the mode
number $\ell$ for different values of $r$. Shown are $f_{r^*}^{\ell}$ for
$r=12M$ (*), $r=6M$ (+), and $r=2.01M$ (x), and $f_{t}^{\ell}$ for
$r=12M$ ($\Diamond$), $r=6M$ (o), and $r=2.01M$ ($\Box$).
}
\label{fig2}
\end{figure}

Three properties of the behavior of the individual modes
are particularly interesting: First, the dipole ($\ell=1$) modes behave
differently than the other modes, and the closer to the BH, the
less important they are. Second, the relative importance of the higher
modes increases approaching the BH. This will require care in the
evaluation of the remainder of the series when we sum over all modes (see
below). Third, as we approach the BH $f_{r^*}^{\ell}\to
f_{t}^{\ell}$. This latter property is obvious from the following
consideration: The covariant components $f^{\ell}_v$ and $f^{\ell}_U$
[where $U$ is the outgoing
Kruskal coordinate, satisfying $U\propto \exp(-u/4M)$ near the EH] assume 
finite values at the EH itself, as $v$ and $U$ are regular coordinates at
the EH. Consequently, $f_{u}^{\ell}$ vanishes exponentially 
with $u$ approaching the EH, yielding 
$f_{t}^{\ell}=f_{v}^{\ell}+f_{u}^{\ell}\to f_{v}^{\ell}$ and
$f_{r^*}^{\ell}=f_{v}^{\ell}-f_{u}^{\ell}\to f_{v}^{\ell}$ as we approach
the EH. Thus, $f_{r^*}^{\ell}\to f_{t}^{\ell}$. 
This is, in fact, a result of spatial gradients becoming comparable to
temporal gradients near the EH. This behavior is shown in
Figure \ref{fig2}(A) for two modes ($\ell=0,1$), but similar behavior is
found also for all the other modes.
Figure \ref{fig2}(B) displays the behavior
of the modes as a function of the mode number $\ell$, for various values
of $r$. The individual modes behave like $\ell^{-2}$ for large values of
$\ell$. Note, that the closer the particle to the BH, the later
the asymptotic $\ell^{-2}$ behavior starts. Most importantly, the detailed
behavior of the modes confirms the expressions for the
analytically-derived regularization parameters \cite{barack-ori,barack}. 

\begin{figure}
\input epsf
\epsfxsize=8.5cm
\centerline{ \epsfxsize 7.5cm
\epsfbox{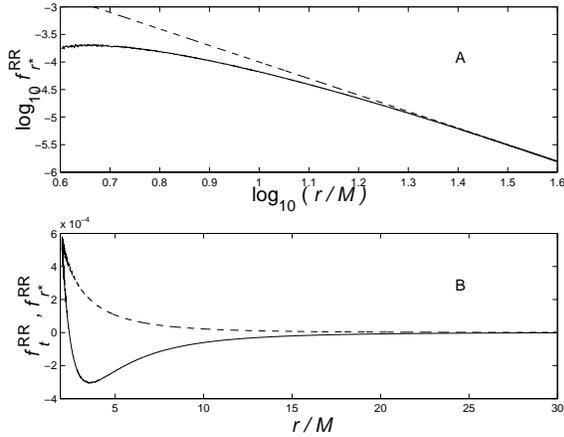}}
\caption{The full RR force as a function of $r$.
Top panel (A): $f_{r^*}^{\rm RR}$ for a marginally-bound worldline (solid
line) and the curve $-0.1\times (r/M)^{-3}$ (dashed line). Bottom panel
(B):
Free fall from rest starting from $r^*_0=40M$.
Dashed line: $f_{t}^{\rm RR}$. Solid line: $f_{r^*}^{\rm RR}$.
}
\label{fig4}
\end{figure}

Next, we sum over all modes to find the total RR force. As noted above,
the relative
importance of the higher modes increases as we approach the horizon. This
causes two problems: (i) it is crucial to include an accurate
approximation of the remainder of the series due to our computation of
only a finite number of modes, and (ii) the noise contribution from the 
$\ell$-mode to the overall force increases with $\ell$. 
The $\ell^{-2}$ behavior of the modes indicates that we can sum over the
modes and calculate the remainder as was done in Ref. \cite{burko-prl}. 
Specifically, the full RR force is 
\begin{equation}
f_{\alpha}^{\rm RR}=\sum_{n=0}^{\ell}f_{\alpha}
^{n\;{\rm  (reg)}}+{\cal R}^{\ell+1}_{\alpha},
\end{equation}
where the remainder can be approximated by
${\cal R}^{\ell+1}_{\mu}\approx \ell^2f_{\mu}^{\ell\;{\rm tail}}
\psi^{(1)}(\ell+1)$. 
Here, $f_{\alpha}^{n\;{\rm  (reg)}}$ is the regularized $\ell$-mode of the
force, and  
$\psi^{(1)}(x)\approx x^{-1}+x^{-2}/2+x^{-3}/6+O(x^{-5})$ is the trigamma 
function. As we sum the
series only up to $\ell=17$, this approximation for ${\cal
R}^{\ell+1}_{\mu}$ guarantees accuracy of $7\times 10^{-4}$ (we neglect
here the contribution to ${\cal R}^{\ell+1}_{\mu}$ from terms 
which scale like $\ell^{-3}$). 
Obviously, approaching the EH $f_{r^*}\to f_{t}$
(as each of the individual modes does). 
Figure \ref{fig4} shows the full RR force
as a function of $r$ for two cases: Fig. \ref{fig4}(A) shows $f_{r^*}^{\rm
RR}$ for a marginally-bound trajectory ($E=1$). At large distances this
force behaves like $f_{r^*}^{\rm RR}\approx -(G/c^2)\beta q^2M/r^3$.
The exponent of $r$ is found here to a $1\%$ accuracy, and we find the
parameter $\beta=(1.00\pm 0.15)\times 10^{-1}$. Figure \ref{fig4}(B) shows
the case of fall from rest, starting from $r^*_0=40M$.  
At large values of $r$ both components of the force
vanish, in accord with the vanishing of the force for a static scalar
charge. The covariant $t$ component, $f_{t}^{\rm RR}$, is everywhere
positive and increases monotonically approaching the BH. This is a
consequence of the particle losing energy by radiating, part of which
escapes to infinity, and the rest being captured by the BH. The
covariant $r^*$ component, $f_{r^*}^{\rm RR}$, is attractive at large
distances. However, near the peak of the effective
potential barrier (near $r^*\approx 0$) its behavior changes, and near
the EH it approaches the value of $f_{t}^{\rm RR}$, as
expected. Note that both components arrive at the EH at a
bounded value. Because $f_{t}^{\rm RR}$ is expected to be positive (the
particle only loses energy by radiating), we infer that $f_{r^*}^{\rm
RR}$ would also be positive approaching the EH, under very general
conditions. In particular, if this behavior persists also for 
charged BHs, and for an electrically-charged particle, then the
properly-defined covariant spatial-component of the RR force at the
EH would be repulsive. If this is indeed the case, then the RR force
acts to reduce the parameter space for which a nearly-extreme spherical
charged BH can be overcharged \cite{hubeny}. 
The question of whether cosmic censorship
for that case is saved by RR effects, however, awaits further
considerations.

We thank Amos Ori, Lee Lindblom, and Kip Thorne for discussions. LMB
wishes to thank the Technion Institute of Theoretical Physics, 
where part of this research was done, for hospitality. At Caltech this
research was supported
by NSF grants AST-9731698 and PHY-9900776 and by NASA grant NAG5-6840.


\end{document}